\pdfoutput=1

\documentclass[superscriptaddress,amsmath,amssymb,twocolumn]{revtex4-1}

\usepackage{graphicx}
\usepackage{chemarr}
\usepackage{times,longtable}
\usepackage{hyperref}
\usepackage{color}
\usepackage{afterpage}
\usepackage{soul,comment}
\usepackage[usenames,dvipsnames]{xcolor}
\usepackage{colortbl}
\usepackage{pbox}
\usepackage{capt-of}
\usepackage{pdfpages}
\makeatletter
\AtBeginDocument{\let\LS@rot\@undefined}
\makeatother

\graphicspath{{large/}}

\bibpunct{[}{]}{,}{n}{}{,}

\newcommand{\f}{\textrm{fer}}
\renewcommand{\r}{\textrm{res}}

\newcommand{\inversefluxunit}{$\text{g}_\text{DW}$h/mmol}

\newcommand{\mmolATPgDW}{$\text{mmol}_\text{ATP}/\text{g}_\text{DW}$}

\usepackage{chemarr}
\usepackage{epsf,mathtools}
\usepackage{graphicx}
\usepackage{dcolumn}
\usepackage{bm}

\begin{document}

\title{\textsf{A yield-cost tradeoff governs Escherichia coli's decision between fermentation and respiration in carbon-limited growth}}

\author{Matteo Mori}
\affiliation{Department of Physics, University of California, San Diego, CA, USA}

\author{Enzo Marinari}
\thanks{Co-last authors}
\affiliation{Dipartimento di Fisica, Sapienza Universit\`a di Roma, Rome, Italy}
\affiliation{INFN, Sezione di Roma 1, Rome, Italy}

\author{Andrea De Martino}
\thanks{Co-last authors}
\affiliation{Soft \& Living Matter Lab, Institute of Nanotechnology (CNR-NANOTEC), Consiglio Nazionale delle Ricerche, Rome, Italy}
\affiliation{Human Genetics Foundation, Turin, Italy}

\begin{abstract}
Many microbial systems are known to actively reshape their proteomes in response to changes in growth conditions induced e.g. by nutritional stress or antibiotics. Part of the re-allocation accounts for the fact that, as the growth rate is limited by targeting specific metabolic activities, cells simply respond by fine-tuning their proteome to invest more resources into the limiting activity (i.e. by synthesizing more proteins devoted to it). However, this is often accompanied by an overall re-organization of metabolism, aimed at improving the growth yield under limitation by re-wiring resource through different pathways. While both effects impact proteome composition, the latter underlies a more complex systemic response to stress. By focusing on {\it E. coli}'s `acetate switch', we use mathematical modeling and a re-analysis of empirical data to show that the transition from a predominantly fermentative to a predominantly respirative metabolism  in carbon-limited growth results from the trade-off between maximizing the growth yield and minimizing its costs in terms of required the proteome share. In particular, {\it E. coli}'s metabolic phenotypes appear to be Pareto-optimal for these objective functions over a broad range of dilutions. 
\end{abstract}

\maketitle

The physiology of cell growth can nowadays be experimentally probed in exponentially growing bacteria both at bulk (see e.g. the bacterial growth laws detailed in \cite{Scott2010interdependence,scott2011bacterial,You2013coordination}) and at single cell resolution \cite{wang2010robust,ullman2013high,jun2015cell,kennard2016individuality}. Refining the picture developed since the 1950s \cite{schaechter1958dependency,kjeldgaard1958}, recent studies have shown that changes in growth conditions are accompanied by a massive re-organization of the cellular proteome, whereby resources are re-distributed among protein classes (e.g. transporters, metabolic enzymes, ribosome-affiliated proteins, etc.) so as to achieve optimal growth performance \cite{hui2015quantitative}. This in turn underlies significant modifications in cellular energetics to cope with the increasing metabolic burden of fast growth \cite{schuetz2007systematic,schuetz2012multidimensional,Basan2015overflow}. 

{\it E.~coli}'s `acetate switch' is a major manifestation of the existence of a complex interplay between metabolism and gene expression. Slowly growing {\it E.~coli} cells tend to operate close to the theoretical limit of maximum biomass yield \cite{ibarra2002evolution,price2004genome}. Fast-growing cells, instead, typically show lower yields, together with the excretion of carbon equivalents such as acetate \cite{Basan2015overflow}. One can argue that, in the latter regime, cells optimize enzyme usage, i.e. they minimize the protein costs associated to growth, while at slow growth they try to use nutrients as efficiently as possible \cite{Basan2015overflow}. As one crosses over from one regime to the other, {\it E. coli}'s growth physiology appears to be  determined to a significant extent by the trade-off between growth and its biosynthetic costs. Interestingly, a similar overflow scenario appears in other cell types in proliferating regimes (see e.g. the Crabtree effect in yeast \cite{de1966crabtree,postma1989enzymic} or the Warburg effect in cancer cells \cite{hsu2008cancer,diaz2009tumor,vander2009understanding}). 

Several phenomenological models have tackled the issue of how metabolism and gene expression coordinate to optimize growth in bacteria \cite{Molenaar2009,scott2011bacterial,flamholz2013glycolytic,maitra2015bacterial}, while mechanistic genome-scale models can provide a more detailed picture of the cross-overs in metabolic strategies that occur as the growth rate is tuned \cite{goelzer2011bacterial,OBrien2013genome,Mori2016constrained}. Here we combine {\it in silico} genome-scale modeling with experimental data analysis to obtain a quantitative characterization of the trade-off between growth and its metabolic costs in {\it E.~coli}. In specific, we show that, for {\it E.~coli} glucose-limited growth, the growth yield (i.e. the growth rate per unit of intaken glucose) is subject to a trade-off with the proteome fraction allocated to metabolic enzymes. At fast (resp. slow) growth, the latter (resp. the former) is optimized, and one crosses over from one scenario to the other as glucose availability is limited. Focusing on energy production and carbon intake as the central limiting factors of growth, we derive an explicit expression linking the biosynthetic costs of growth to the growth rate and evaluate it within a genome-scale model of {\it E.~coli}'s metabolism where the costs associated to different strategies for ATP synthesis can be directly assessed. The ensuing trade-off is described in terms of a Pareto front in a two-objective function landscape. Remarkably, {\it E. coli}'s metabolic phenotypes are found to be Pareto-optimal over a broad range of growth rates, a picture that we validate through an analysis of proteomic data. This study therefore provides a quantitative characterization of the multi-dimensional optimality of living cells \cite{schuetz2012multidimensional,hart2015inferring} that directly addresses the crosstalk between growth physiology and gene expression.

\section*{Results}

\subsection*{General view of the optimal proteome allocation problem} \label{sect:general_model}

Because growth is severely affected by the synthesis of inefficient proteins \cite{Scott2010interdependence}, optimizing proteome composition is a major fitness strategy for exponentially growing bacteria in any given condition. In {\it E.~coli}, for instance, a substantial reshaping of the proteome takes place in carbon-limited growth, with ribosome-affiliated proteins taking up an increasing fraction of the proteome as the bulk growth rate $\mu$ increases, at the expense of catabolic, motility and biosynthetic proteins  \cite{Scott2010interdependence,hui2015quantitative,schmidt2016quantitative}. Similar changes are observed in cells subject to other modes of growth limitation, like nitrogen starvation or  translational inhibition \cite{hui2015quantitative}. 

\begin{figure*}
 \centering
   \includegraphics[width=\textwidth]{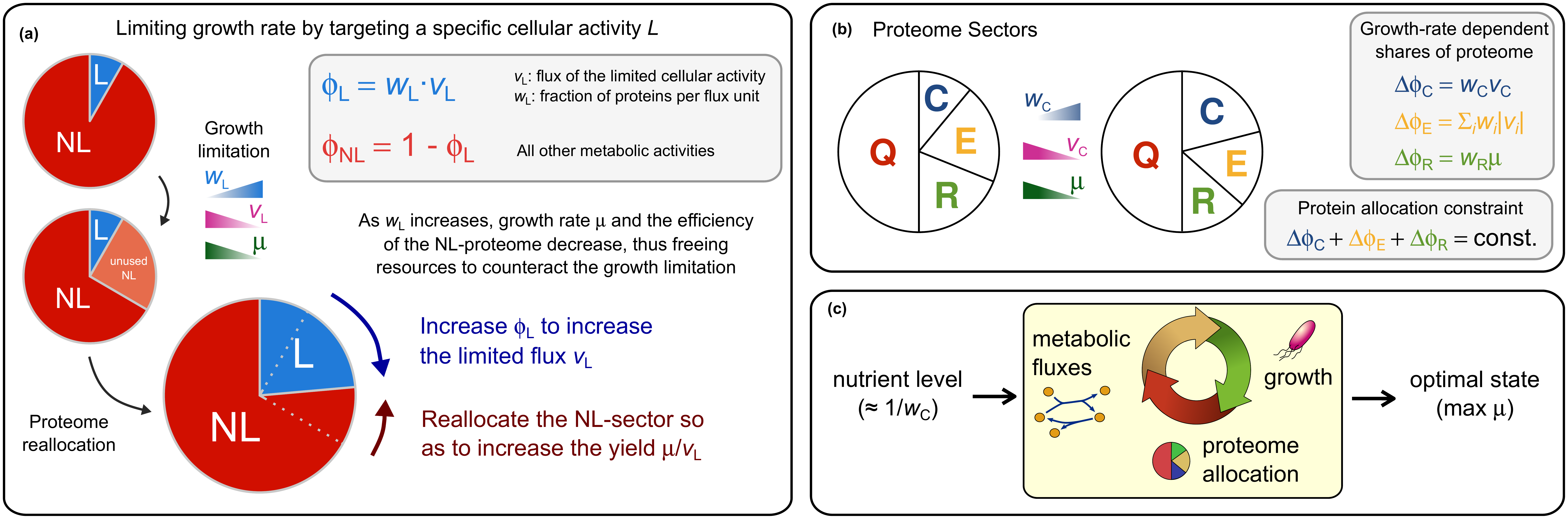}
 \caption{{\bf Schematic view of the proteome allocation problem.} (a) Effect of growth limitation on proteome composition. An increase in the cost of the limiting sector can be counteracted by expanding the limiting sector (i.e. by increasing $\phi_L$) and/or by allocating part of the unnecessary non-limiting sector to re-organizing metabolism so as to increase the growth yield $Y=\mu/v_L$. (b) Proteome sectors considered for {\it E.~coli}, namely ribosomal (R), enzymatic (E), catabolic (C), and core (Q). Following \cite{Mori2016constrained}, we assume that all sectors but the core have $\mu$-dependent parts $\Delta\phi_j$. By normalization, their sums are constrained as in Eq.~\eqref{eq:CAFBA_constraint2}. (c) A change in the nutrient level is processed by the cell via metabolic fluxes, which affect growth. As the cell senses its new state, it re-allocates its proteome, thereby modulating metabolic fluxes and improving growth. The interplay between the various components leads to optimal phenotypes.}
\label{fig:piechart_schemes} 
\end{figure*}

In general terms, the problem of optimal proteome allocation can be posed as follows. Consider a generic cellular activity $L$ described by a rate $v_L$ that is subject to limitation and such that $\mu$ is proportional to $v_L$ via a `yield' $Y$ representing the growth rate per unit of $v_L$, so that $\mu=Y v_L$. For instance, $v_L$ may be the rate at which a nutrient is imported from the growth medium and metabolized, which is limited e.g. by nutrient availability; or the rate of an intracellular flux reduced by specific stresses (e.g. high levels of toxic metabolites or antibiotics). In this scenario, as the stress is applied, $v_L$ decreases and $\mu$ is proportionally reduced. Generically, more $L$-devoted proteins will be needed to sustain a given rate $v_L$ under stress. We denote the proteome share allocated to $L$-devoted proteins by $\phi_L$ (with the rest of the proteins sizing up to a fraction $\phi_{NL}=1-\phi_L$ of the total), and define the  ``proteome cost'' $w_L$ of sustaining rate $v_L$ via $\phi_L=w_L v_L$. In these terms, the growth rate $\mu$ is proportional to the growth yield $Y$ and to $\phi_L$, and inversely proportional to $w_L$, i.e.
\begin{equation}
\mu=\frac{\phi_L \cdot Y}{w_L}~~.
\end{equation}
This expression shows that cells can counteract an increase of $w_L$, i.e. a stress affecting $L$, in two ways, see Fig.~\ref{fig:piechart_schemes}a.
\begin{enumerate}
\item The first is by increasing $\phi_L$. This strategy underlies e.g. an upregulated synthesis of transporters and catabolic proteins in response to a nutrient shortage, or an increase of the ribosomal proteome fraction in response to antibiotics, as seen for instance in \cite{hui2015quantitative}. 
\item In addition, they can try to increase the growth yield $Y$ (or, equivalently, decrease of the specific flux through the limited process, i.e. $q\equiv Y^{-1}=v_L/\mu$). The growth yield is however a systemic property that depends on the whole set of metabolic processes. Achieving a more efficient conversion of $v_L$ to $\mu$ therefore requires a re-organization of the entire non-limiting sector that occupies a fraction $\phi_{NL}$ of the proteome. 
\end{enumerate}
{\it E.~coli}'s `acetate switch' \cite{wolfe2005acetateSwitch,Basan2015overflow}, whereby bacteria cross over from a predominantly fermentative to a predominantly respiratory metabolism upon carbon limitation, is an example of the latter strategy. Indeed respiration, while more costly in terms of enzymes with respect to fermentation, has a larger ATP yield (ca. 26 mol$_{\rm ATP}$/mol$_{\rm glc}$ versus ca. 12 mol$_{\rm ATP}$/mol$_{\rm glc}$ \cite{Basan2015overflow}). 

Cells may employ combinations of the above strategies, as for instance {\it E.~coli} in glucose limitation tends to both increase the fraction of proteins devoted to glucose scavenging and import, and to switch to the more efficient respiratory pathways. However, the metabolic re-wiring required to increase $Y$ necessarily implies the coordinated modulation of expression levels across multiple metabolic pathways. In the following we will aim at characterizing more precisely the trade-off that underlies such a re-organization.

\subsection*{Proteome sectors in {\it E.~coli}}

A consistent body of experimental work has shown that, in exponential growth, {\it E.~coli}'s proteome can be partitioned into ``sectors'' whose relative weights adjust with the growth conditions \cite{Scott2010interdependence,You2013coordination,hui2015quantitative}. At the simplest level, a four-way partition can be considered, in which three sectors [ribosome-affiliated proteins ($R$), metabolic enzymes ($E$) and proteins involved in the uptake system of the limiting nutrient ($C$)] respond to the growth rate $\mu$, while a fourth sector (a core $Q$ formed by housekeeping proteins) is $\mu$-independent. Normalization of proteome mass fractions imposes 
\begin{equation}\label{eq:CAFBA_constraint_1}
\phi_C(\mu)+\phi_R(\mu)+\phi_E(\mu)=\underbrace{1-\phi_Q}_{\text{$\mu$-independent}}~~.
\end{equation}
Based on the bacterial `growth laws' characterized in \cite{Scott2010interdependence,hui2015quantitative}, each $\mu$-dependent term in (\ref{eq:CAFBA_constraint_1}) can take the form $\phi_X(\mu)=\phi_{X,0}+\Delta\phi_X(\mu)$, with $\phi_{X,0}$ an offset value and $X\in\{R,E,C\}$. In turn, metabolic fluxes can be seen as the brokers of proteome re-shaping assuming they are proportional to enzyme levels \cite{Scott2010interdependence,You2013coordination,Mori2016constrained}. In particular, for carbon-limited growth (\ref{eq:CAFBA_constraint_1}) can be re-cast as (see Fig.\ref{fig:piechart_schemes}b, \cite{You2013coordination,Mori2016constrained})
\begin{equation} \label{eq:CAFBA_constraint2}
\underbrace{w_C v_C}_{\Delta\phi_C} + \underbrace{\sum_{i\in E} w_i |v_i|}_{\Delta\phi_E} + \underbrace{w_R \mu}_{\Delta\phi_R} = \underbrace{\phi_{\max}}_{\text{$\mu$-independent}}~~,
\end{equation}
where $v_C$ is the rate of carbon intake, $v_i$ is the flux of reaction $i$, the sum runs over enzyme-catalyzed reactions and $\phi_{\max}$ is a constant that includes all $\mu$-independent terms (equal to about $0.48$ or 48\% in {\it E.~coli}). The three terms on the left-hand side of (\ref{eq:CAFBA_constraint2}) give explicit representations to the $\mu$-dependent part of the proteome. The first term corresponds to the $\mu$-dependent proteome fraction to be allocated to the $C$-sector, with $w_C$ the cost of sustaining a carbon intake flux $v_C$ (i.e. the proteome share to be allocated to $C$ per unit of carbon influx). As detailed in \cite{Mori2016constrained}, $w_C$ reflects the amount of carbon available in the extracellular medium, with large values (high import costs) associated to low carbon levels. The term $w_R\mu$ describes the empirically observed linear increase of $\phi_R$ with the growth rate \cite{Scott2010interdependence,You2013coordination,hui2015quantitative}, the coefficient $w_R$ corresponding to the proteome fraction to be allocated to the $R$-sector per unit of $\mu$ (in short, the ``cost'' of $R$). In {\it E.~coli}, $w_R$ is determined in a robust way by regulation of ribosome expression via ppGpp \cite{potrykus2008p}, which sets its value to the inverse translational capacity $w_R\simeq 0.169$~h \cite{Mori2016constrained}. The last term represents instead the $\mu$-dependent part of the $E$-sector. The coefficients $w_i$ quantify the cost of each reaction $i\in E$ in terms of the proteome fraction to be allocated to its enzyme per unit of net flux. For sakes of simplicity, we assume here that $w_i$ is the same for each $i$ (but see \cite{Mori2016constrained} for a discussion of alternative choices). 

Note that (\ref{eq:CAFBA_constraint_1}) and (\ref{eq:CAFBA_constraint2}) have the form $w_L v_L + \phi_{NL}=1$ upon identifying $w_L v_L$ with $\Delta\phi_C\equiv w_C v_C$ and including the $\mu$-dependent $E$- and $R$-sectors into $\phi_{NL}$. Coherently with the general problem of protein allocation, one expects that $\phi_{NL}$, and hence $\Delta \phi_E$, might re-shape to counteract nutrient limitation. Crucially, though, $\Delta\phi_E$ depends on intracellular fluxes and takes on different values in different metabolic states, so that, in principle, any re-shaping of the non-limiting sector will be tied to a change in the overall organization of metabolic activity. Therefore, nutrient stress (i.e. an increase of the nutrient import cost $w_C$) will be mediated by metabolic fluxes into a re-organization of the cellular proteome that in turn affects $\mu$ (see Fig. \ref{fig:piechart_schemes}c). Optimal growth at each nutrient level (i.e. at each $w_C$) results from the crosstalk between proteome allocation and metabolism.

\subsection*{Optimal proteome fractions are constrained within tight bounds and interpolate between them as the growth rate changes} \label{sect:review}

\begin{figure*}
 \centering
   \includegraphics[width=\textwidth]{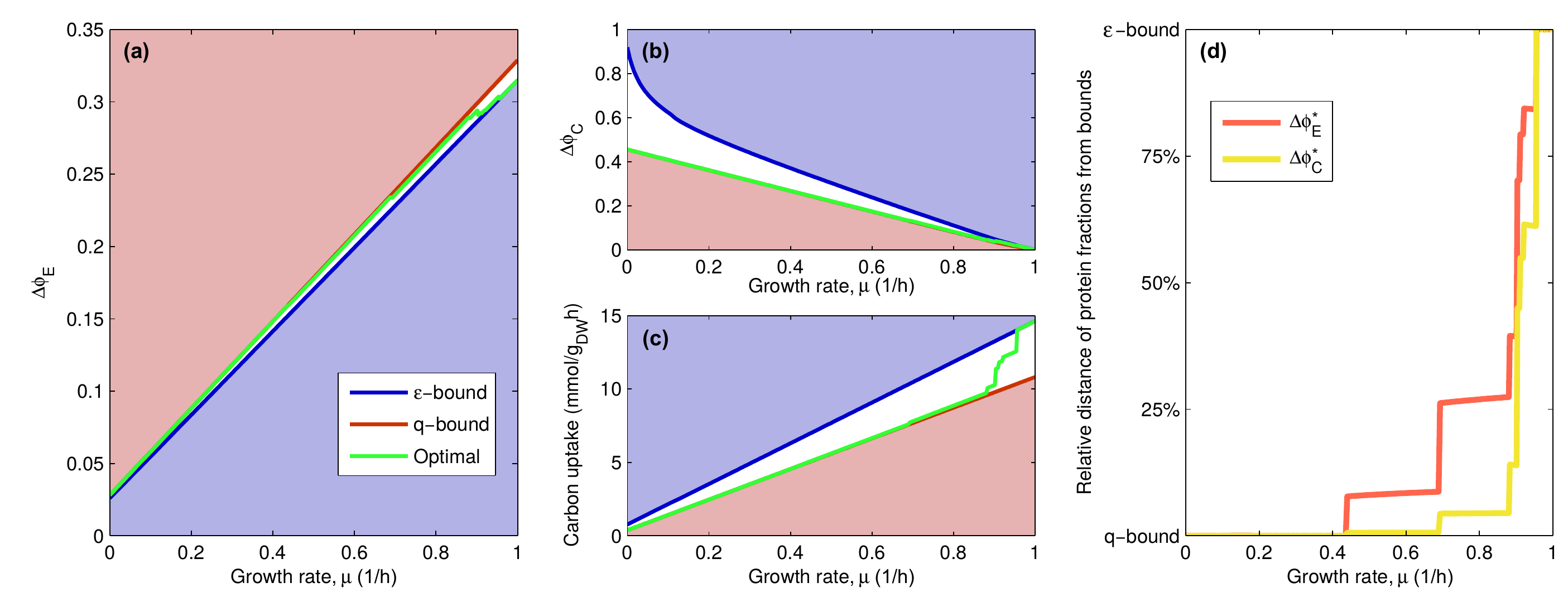}
 \caption[Feasibility regions for different growth rates $\mu$.]  {\small {\bf Allowed ranges of variability and optimal values for proteome fractions}. Feasibility regions for $\Delta\phi_E$ ({\bf a}), $\Delta\phi_C$ ({\bf b}) and carbon uptake $v_C$ ({\bf c}), as functions of the growth rate $\mu$. For any $\mu$, the $q$- and the $\varepsilon$-bound are computed by minimizing either the carbon flux $v_C$ or the E-sector proteome share $\Delta\phi_E$. Optimal, $\mu$-maximizing proteome fractions, represented by green lines, interpolate between these bounds as $\mu$ changes. ({\bf d}) Fractional distance of the optimal $C$- and $E$-sector fractions, $\Delta\phi_C^\star$ and $\Delta \phi_E^\star$, to the $q$- and $\varepsilon$-bounds. For slow growth, both $\Delta\phi_C^\star$ and $\Delta \phi_E^\star$ are close to the $q$-bound. As $\mu$ increases, they both shift toward the $\varepsilon$-bound. Note that $\Delta\phi_E^{(\varepsilon)} \le \Delta\phi_E^\star \le \Delta\phi_E^{(q)}$ while $\Delta\phi_C^{(q)} \le \Delta\phi_C^\star \le \Delta\phi_C^{(\varepsilon)}$.}
 \label{fig:2Dregions} 
\end{figure*}

In general, any flux pattern $\mathbf{v}=\{v_C,\{v_i\}_{i\in E}\}$ compatible with (\ref{eq:CAFBA_constraint2}) and with minimal mass balance conditions is a viable non-equilibrium steady state for {\it E.~coli}'s metabolic network. Optimal flux patterns, and hence optimal values of $\Delta\phi_E=\sum_{i\in E}w_i|v_i|$ and $\Delta\phi_C=w_C v_C$ (which we denote respectively as $\Delta\phi_E^\star$ and $\Delta\phi_C^\star=w_C v_C^\star$), correspond to maximum $\mu$. In order to calculate such $\mu$-maximizing flux patterns $\mathbf{v}^\star$, we resorted to genome-scale constraint-based modeling (see Methods). Focusing on {\it E.~coli} in a glucose-limited minimal medium, we obtained the green curves shown in Fig. \ref{fig:2Dregions}a--c, detailing how these quantities vary with the growth rate $\mu$. Quite surprisingly, both $\Delta\phi_E^\star$ (Fig.~\ref{fig:2Dregions}a) and $\Delta\phi_C^\star$ (Fig.~\ref{fig:2Dregions}b) appear to display an almost linear behaviour with $\mu$. Significant deviations occur only at fast growth, more evidently for $\Delta\phi_E^\star$ and for $v_C^\star$, see Fig.~\ref{fig:2Dregions}c. 

This behaviour suggests that optimal proteome fractions are tightly constrained by either optimality or proteome allocation requirements. The allowed ranges of variability of $\Delta\phi_E$ and $\Delta\phi_C$ irrespective of (\ref{eq:CAFBA_constraint2}) can be computed {\it in silico} by Linear Programming (LP, see Methods and Supporting Text for mathematical details). We call these the $q$-bound (index $(q)$) and the $\varepsilon$-bound (index $(\varepsilon)$), respectively, such that
\begin{gather}
 \Delta\phi_C^{(q)} \le \Delta\phi_C^\star \le \Delta\phi_C^{(\varepsilon)} \label{eq:bound1}\\
 \Delta\phi_E^{(\varepsilon)} \le \Delta\phi_E^\star \le \Delta\phi_E^{(q)} \label{eq:bound2}
\end{gather}
(Note that the $q$-bound is above the $\varepsilon$-bound for $\Delta\phi_E$, and vice-versa for $\Delta\phi_C$, i.e. when one of the two quantities is minimal the other should stick to its largest allowed value.) The bounds corresponding to {\it E. coli} growth in glucose-limited minimal medium are reported in Fig. \ref{fig:2Dregions}a--c as red and blue curves, respectively. 

One clearly sees that $\Delta\phi_E^\star$ nearly saturates its maximum (given the $q$-bound) for slow growth and gradually shifts to its minimum (given the $\varepsilon$-bound) as $\mu$ increases. Vice-versa, $\Delta\phi_C^\star$ interpolates between its minimum (given the $q$-bound) and its maximum (given the $\varepsilon$-bound) as growth gets faster. This is clearly visible in Fig. \ref{fig:2Dregions}d, which clarifies how close $\Delta\phi_C$ and $\Delta\phi_E$ are to their respective bounds. For slow growth  (below ca. 0.5/h), both the $E$- and $C$-sector saturate their $q$-bounds and shift discontinuously to the $\varepsilon$-bound at higher $\mu$. 

These results suggest that phenotypes minimizing nutrient import costs (in terms of proteome shares) are optimal at slow growth, whereas phenotypes minimizing enzyme costs are favored at fast growth. We stress that the $q$- and $\varepsilon$-bounds do not account for (\ref{eq:CAFBA_constraint2}), implying that the optimal proteome allocation compatible with (\ref{eq:CAFBA_constraint2}) interpolates between the physically allowed limits. In the broad cross-over region cells appear to balance between the costs of importing glucose and those of metabolic processing. Optimality therefore appears to be  generically characterized by a trade-off between different cost functions, with  ``extreme'' conditions (e.g. fast versus slow growth) favoring the minimization of one over the other.  In between, the trade-off is strongest and the cell has to fine-tune its metabolism so as to optimally balance the two objectives. Such a tradeoff can be shown to occur under very general assumptions, i.e. without the need of specifying a detailed functional form for the protein sectors in terms of the fluxes (see Supporting Text). In other terms, as long as the growth rate is maximized, an increase of the growth yield has be accompanied by an increase of the protein cost of metabolism, encoded in the the non-limiting proteome sector (see Fig. \ref{fig:piechart_schemes}a).  

\subsection*{Growth yield and enzyme costs are subject to a trade-off at optimal growth} \label{sect:pareto}
 
The above scenario can be re-cast in more intuitive terms as follows. Let us assume that each metabolic flux $v_i$ scales proportionally to the growth rate $\mu$, i.e. that $\mathbf{v} = \boldsymbol{\xi} \cdot \mu$, with $\boldsymbol{\xi}$ a representative flux vector identifying the ``metabolic state'' of the cell. (While this approximation is made for theoretical convenience here, it is empirically valid for moderate to fast growth rates \cite{neijssel1996growth}.) We may now isolate $\mu$ from (\ref{eq:CAFBA_constraint2}), obtaining 
\begin{equation}\label{eq:mu_function_of_q_epsi}
\mu(\boldsymbol{\xi})=\frac{\phi_{\max}}{w_R+\varepsilon(\boldsymbol{\xi})+w_C q(\boldsymbol{\xi})}~~,
\end{equation}
where $\varepsilon\equiv\Delta\phi_E/\mu = \sum_i w_i |\xi_i|$ stands for the specific cost of the $E$-sector, whereas $q\equiv v_C/\mu=\xi_C$ denotes the specific carbon uptake (i.e. the amount of in-taken carbon per unit of growth rate, or the inverse growth yield $Y^{-1}$). Because in {\it E.~coli} $w_R$ and $\phi_{\max}$ are roughly constant, (\ref{eq:mu_function_of_q_epsi}) relates directly the growth rate $\mu(\boldsymbol{\xi})$ of a given flux pattern $\boldsymbol{\xi}$ to its overall ``specific cost'' 
\begin{equation}
\mathcal{C}(\boldsymbol{\xi})=\varepsilon(\boldsymbol{\xi})+w_C q(\boldsymbol{\xi})~~. 
\end{equation}
More specifically, one sees that maximizing $\mu$ is equivalent to minimizing $\mathcal{C}$ across the different metabolic states $\boldsymbol{\xi}$. Ideally, at optimality cells would like to minimize $\mathcal{C}$ by minimizing $\varepsilon$ and $q$ independently. However both quantities depend on the underlying metabolic state $\boldsymbol{\xi}$, so that, as $w_C$ varies, optimal states must strike a compromise between growth yield and biosynthetic costs. In particular, for $w_C\to 0$ (i.e. in carbon-rich media), $\mu$ is maximized by minimizing $\varepsilon$, a scenario that corresponds to the $\varepsilon$-bound described in the previous section, that optimal metabolic flux patterns saturate at fast growth. On the other hand, when $w_C\gg 1$ (i.e. when extracellular glucose levels are low), $\mu$ is maximized by maximizing the growth yield $Y$ (or by minimizing $q$), leading to the $q$-bound that is saturated at slow growth at optimality. Intermediate glucose levels require instead a trade-off between these two objectives. Notice that this scenario is fully consistent with the fermentation-to-respiration switch that characterizes {\it E. coli} growth in carbon limitation and with the idea that its metabolism is multi-objective optimal \cite{schuetz2012multidimensional}.

\subsection*{Quantifying the yield-cost trade-off: Pareto-optimality of {\it E. coli}'s metabolism}

\begin{figure*}
 \centering
 \includegraphics[width=\textwidth]{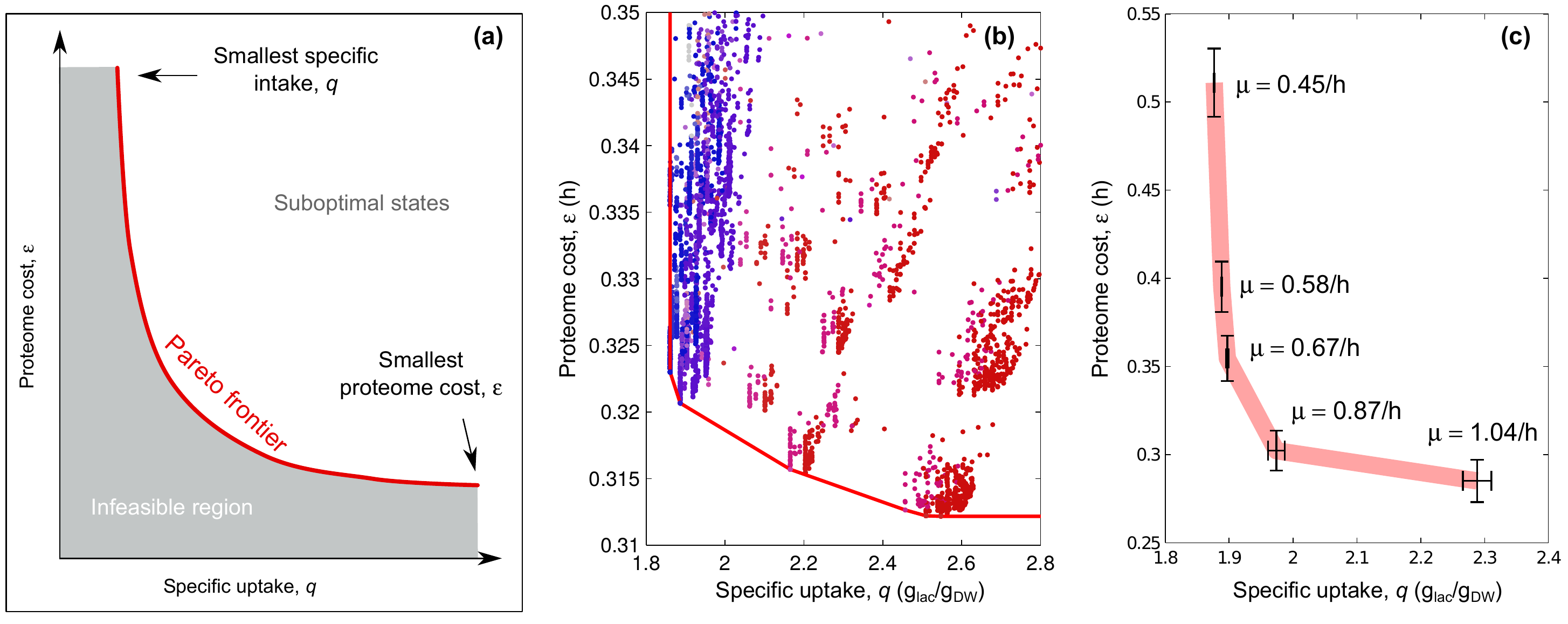}
 \caption{\label{fig:ecoli_pareto} {\bf Trade-off between maximum yield and minimum enzyme cost in {\it E. coli}.}
(a) Multi-objective optimality and Pareto front. Two cost functions (specific carbon intake $q$ and specific proteome cost $\varepsilon$) are shown, together with the feasibile (white) and infeasible (grey) regions, separated by the Pareto frontier. Optimal solutions lie on the latter. (b) {\it In silico} prediction for optimal {\it E.~coli} growth on lactose-limited  minimal medium. The red line corresponds to the computed Pareto front (see Methods), while individual points in the feasible region describe sub-optimal solutions. Blue (resp. red) markers represent solutions dominated by respiration (resp. fermentation), while purple markers denote mixtures. (c) {\it E. coli} states obtained by integrating mass spectroscopy data for lactose-limited growth from \cite{hui2015quantitative} with {\it in silico} predictions qualitatively reproduce (with quantitative accuracy for the yield) the predicted Pareto front. The values of $\mu$ reported next to the experimental points represent the experimental growth rates.} 
\end{figure*}

In the present context, multi-objective optimality can be described quantitatively by a Pareto frontier that separates an accessible region of the $(q,\varepsilon)$ plane, such that each point lying therein corresponds to a viable metabolic phenotype, from an inaccessible one, with optimal states lying on the front (see Fig.~\ref{fig:ecoli_pareto}a). Fig.~\ref{fig:ecoli_pareto}b shows the Pareto frontier of optimal metabolic phenotypes we obtained for lactose-limited {\it E.~coli} growth using constraint-based modeling (see Methods and Supporting Text).  The growth rate $\mu$ increases as one moves along the Pareto front towards larger values of $q$ (i.e. lower yields). Sub-optimal states, generated by a randomized constraint-based model (see Methods), lie as expected in the feasible region. Both optimal and sub-optimal flux patterns show a robust switch to a low-yield phenotype at fast growth rates, characterized by acetate secretion and downregulated respiration \cite{Mori2016constrained}. Such solutions dominate at large values of the inverse yield $q$. Instead, fluxes through the TCA cycle and the glyoxylate shunt are mostly active in the high-yield flux patterns that mainly characterize slow growth. 

We have validated {\it E. coli}'s Pareto-optimality scenario against experimental results for lactose-limited {\it E.~coli} growth by first computing $\varepsilon$ from mass spectrometry data, and then by assigning a growth yield to each state thus obtained by further constraining {\it in silico} models with the empirically found values of $\varepsilon$ (see Methods). This yields the curve in the $(q,\varepsilon)$ plane shown in Fig.~\ref{fig:ecoli_pareto}c, which displays a remarkable qualitative agreement with our computation, confirming the cost-yield trade-off scenario. At the quantitative level, we note that the normalized protein cost $\varepsilon$ predicted {\it in silico} for high-yield states matches the observed enzyme cost at growth rate $\mu= 0.67/$h. For faster rates, where acetate excretion sets in, our model underestimates the decrease in $\varepsilon$ by only about 10\%. Likewise, at slow growth (below 0.6/h), our prediction appears to underestimate $\varepsilon$, most likely due to the decrease in enzyme efficiencies that is known to set in at low $\mu$ \cite{Bennett2008absolute, Boer2010growth, Valgepea2013escherichia, OBrien2013genome, OBrien2016quantification} and which is not accounted for in the constraint-based framework we employed.

\section*{Discussion} \label{sect:discussion}

\subsection*{{\it E.~coli}'s acetate switch as a two-state system} \label{sect:coarse_grained}

\begin{figure*}
 \centering
 \includegraphics[width=17cm]{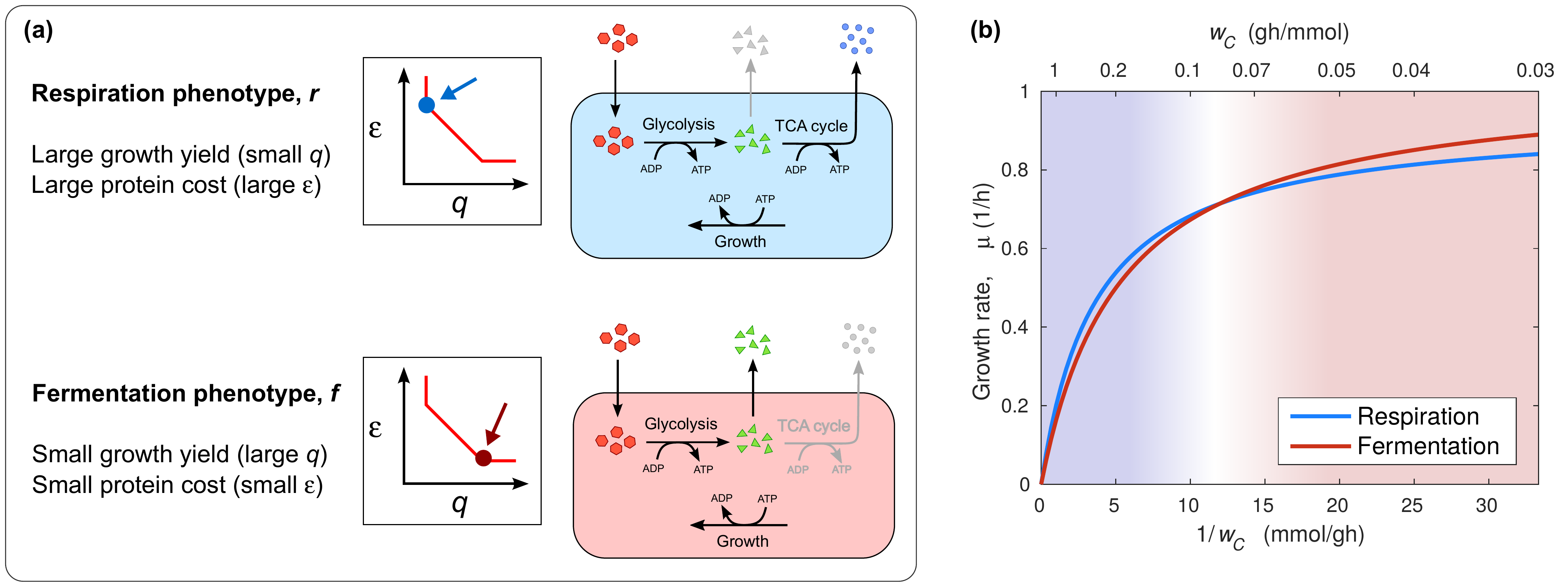}
 \caption[FBA and CAFBA solutions of the coarse grained model.]{\small {\bf Phenomenological two-state view of {\it E. coli} carbon-limited growth.} (a) Respiration and fermentation phenotypes as characterized by the multi-objective optimal states on the Pareto frontier of {\it E.~coli}'s metabolism. The respiration phenotype has a large yield (small $q$) and large specific protein costs, while the fermentation phenotype carries lower yields (higher $q$) and a smaller cost. (b) Growth rate ($\mu$) versus carbon-intake cost $w_C$ as obtained from the phenomenological two-state model discussed in the Supporting Text. For each $w_C$, the optimal phenotype is the one for which $\mu$ is largest. The switch from the fermentation to the respiration phenotype occurs when $w_C$ matches the extra protein cost required by respiration.} 
 \label{fig:coarse_grained_model_solutions}
\end{figure*}

In summary, our results indicate that, in the case of {\it E.~coli}, the range of values of $w_C$ where the yield-cost trade-off is significant is relatively small. It is therefore reasonable to classify flux patterns on the Pareto frontier in two broad types (see Fig. \ref{fig:coarse_grained_model_solutions}a). The first one corresponds to a `fermentation' phenotype with low yields ($q_\f\gtrsim 2.3$ q$_{\rm lac}/$q$_{\rm DW}$) but low specific protein cost ($\varepsilon_\f\simeq 0.3$ h), characterized by carbon overflow and robust flux through fermentative pathways. The second one has higher yield ($q_\r \simeq 1.9$ q$_{\rm lac}/$q$_{\rm DW}$) but higher costs ($\varepsilon_\r\gtrsim 0.35$ h), and uses respiration as its major energy-producing pathway. Generic flux patterns can be seen as linear combinations of these phenotypes with parameter $\alpha$ ($0\leq\alpha\leq 1$), giving inverse yield $q(\alpha)=\alpha q_\r+(1-\alpha)q_\f$ and carrying a cost $\varepsilon(\alpha)=\alpha \varepsilon_\r+(1-\alpha)\varepsilon_\f$. Correspondingly, the growth rate $\mu$ can be computed as a function of $\alpha$ from (\ref{eq:mu_function_of_q_epsi}), i.e. 
\begin{equation}
\mu(\alpha)=\frac{\phi_{\max}}{w_R+\varepsilon (\alpha)+w_C q(\alpha)}~~.
\end{equation}
One can see (see Supporting Text) that $\mu$ is maximized by the respiration phenotype with $\alpha=1$ (resp. the fermentation phenotype with $\alpha=0$) when $w_C$ is above (resp. below) the value 
\begin{equation}
w_C^{\rm ac} \equiv \frac{\varepsilon_\r-\varepsilon_\f}{q_\f-q_\r}\simeq 0.1~\mathrm{h}~~,
\end{equation} 
corresponding to a growth rate $\mu_{\rm ac} \simeq 0.7$/h, in quantitative agreement (within 10\%) with the experimentally determined onset of the acetate switch \cite{Basan2015overflow}. 

Such a two-state scenario inspires a minimal coarse-grained mathematical model of {\it E.~coli}'s metabolism in which the cell can use either respiration or fermentation to produce energy subject to a global constraint on proteome composition (see Supporting Text). The model predicts that, at optimality, a transition between the fermentation phenotype (fast growth) and the respiration phenotype (slow growth) occurs when the cost of  intaking carbon matches the extra protein cost required by respiration, at which point one phenotype outperforms the other in terms of maximum achievable growth rate (see Fig. \ref{fig:coarse_grained_model_solutions}b).  Straightforward mathematical analysis furthermore helps clarifying how constraints associated to proteome costs differ from other types of mechanisms that have been suggested to drive the acetate switch in {\it E. coli} (see Supporting Text).

\subsection*{Outlook}

The Pareto scenario presented above allows to describe,  with quantitative accuracy, the complex cellular economics underlying {\it E.~coli} growth in carbon-limited media in terms of a multi-objective optimization problem, and ultimately leads to a minimal, two-state model of {\it E.~coli}'s metabolism that includes its essential features. Many coarse grained models of the switch between respiration and fermentation are, in fact, two-state models of the kind we have described \cite{Molenaar2009, Basan2015overflow,vazquez2016macromolecular}. The very recent model of Basan {\it et al}. \cite{Basan2015overflow}, in particular, addresses specifically the impact of protein costs on the emergence of fermentation metabolism. The approach employed here differs in two points. First, the yields and proteome cost parameters for respiration and fermentation used in \cite{Basan2015overflow} refer to the ATP yield (as opposed to the growth yield) and to specific  ``respiration'' and ``fermentation'' proteomes. In the present model, both pathways are part of the same $E$-sector, and the focus is on a global re-allocation of the proteome rather than on up- or down-regulation of specific pathways. Secondly, and more importantly, the cost of carbon uptake (i.e. $w_C$) is implicitly assumed to be nil in \cite{Basan2015overflow}. When $w_C$ is set to zero, metabolism is completely determined by the normalization of proteome fractions and by the energy flux balance. While the switch to fermentation is still a consequence of proteome allocation, its physical origin is rather different in the two models. In \cite{Basan2015overflow}, it is enforced by the energy demand. Under Pareto optimality, instead, it is a consequence of the tradeoff between the $C$- and $E$-sectors. This specific aspect makes it in principle possible to describe strains with different ``acetate overflow lines'' (e.g. mutants \cite{castano2009insight,Valgepea2010systems} or ``acetate feeding'' strains obtained in evolution experiments \cite{helling1987evolution,treves1998repeated}), which correspond to feasible --albeit suboptimal-- cellular states that would be harder to describe by the model of \cite{Basan2015overflow}. On the other hand, the latter characterizes, in a sense, an ``optimal'' strain. Future experiments may allow to measure fitness advantages of different  metabolic strategies in various environmental and ecological contexts, shedding further light on the evolution of the acetate switch.

By slightly extending the model of \cite{Basan2015overflow}, Vazquez and Oltvai \cite{vazquez2016macromolecular} have recently linked overflow metabolism to a macromolecular crowding constraint, along the lines of \cite{Beg2007,vazquez2008impact}. For {\it E.~coli}, such an interpretation appears to be at odds with the empirical fact that the cell volume adjusts in response to changes in the macromolecular composition of the cell. In particular, the cell density was found to be roughly constant across several distinct conditions, including inhibition of protein synthesis, and only slightly larger in the case of protein over-expression \cite{woldringh1981variation,Basan2015inflating}. The fact that cell density is minimally perturbed by ``inflating'' or ``deflating'' cells via tuning of protein synthesis suggests a reduced role of macromolecular crowding in modulating such processes in {\it E. coli}. In addition, however,  \cite{vazquez2016macromolecular} points out that, at slow growth,    an increase of the proteome share of proteins other than those associated to respiration and fermentation has to take place. Our results are in line with this scenario. In fact, catabolic proteins included in the $C$-sector are up-regulated at low $\mu$ (see Fig. \ref{fig:2Dregions}b), in agreement with quantitative measurements \cite{hui2015quantitative}. It is indeed the relationship between the $C$- and $E$-sectors, the latter of which accounts for respiration and fermentation pathways, that we have focused on in this work. 

It is known that many different organisms share {\it E.~coli}'s behaviour in terms e.g. of growth laws \cite{scott2011bacterial} and carbon overflow. Nevertheless, the picture derived here for {\it E.~coli} is not universally valid across microbial species. For instance, recent studies of {\it L. lactis}, an industrial bacterium that displays carbon overflow (albeit between different types of fermentation pathways rather than between fermentation and respiration as {\it E.~coli}), suggest that protein costs are not a determinant factor in its growth strategies \cite{goel2015protein}. Likewise, carbon overflow in {\it S. cerevisi\ae} appears to respond to the glucose intake flux rather than to the macroscopic growth rate \cite{huberts2012flux}. More work is therefore required to clarify the extent to which the picture described here applies to other organisms.

\section*{Methods} \label{sect:materials_methods}

\subsection*{Metabolic network reconstruction}

All computations were performed on {\it E. coli}'s iJR904 GSM/GPR genome-scale metabolic model \cite{Reed2003} using a glucose-limited or a lactose-limited minimal medium.

\subsection*{Computation of the $q$- and $\varepsilon$-bounds and of the optimal values of $\Delta\phi_C$ and $\Delta\Phi_E$ via constraint-based modeling}

Flux Balance Analysis (FBA \cite{orth2010flux}) approaches to metabolic network modeling search for optimal flux vectors $\mathbf{v}=\{v_i\}$ within the space $\mathcal{F}$ defined by the mass balance conditions $\mathbf{Sv=0}$, $\mathbf{S}$ denoting the stoichiometric matrix, and by thermodynamic constraints imposing that $v_i\ge 0$ for irreversible reactions. The $q$- and $\epsilon$-bounds are obtained by solving
\begin{align}\label{r}
\text{$q$-bound :}& ~~:~~ \displaystyle\min_{\mathbf{v}\in\mathcal{F}} \, \Delta\phi_C ~~ \text{subject to $\mu(\mathbf{v})=\mu_0$}\\
\label{f}
\text{$\varepsilon$-bound:} & ~~:~~ \displaystyle\min_{\mathbf{v}\in\mathcal{F}} \,\Delta\phi_E ~~ \text{subject to $\mu(\mathbf{v})=\mu_0$}~~,
\end{align} 
upon varying $\mu_0$, where $\mu(\mathbf{v})$ denotes the growth rate associated to $\mathbf{v}$. Both problems are solved by LP and we employed the openCOBRA toolbox \cite{schellenberger2011quantitative} for their solution. The growth-rate dependent minimum values attained by the objective functions, which we denoted by $\Delta\phi_C^{(q)}$ and $\Delta\phi_E^{(\varepsilon)}$ in (\ref{eq:bound1}) and (\ref{eq:bound2}) respectively, directly provide the lower bounds for $\Delta\phi_C$ and $\Delta\phi_E$. The upper bounds  $\Delta\phi_E^{(q)}$ and $\Delta\phi_C^{(\varepsilon)}$ can be computed from the flux vectors $\mathbf{v}^{(q)}$ and $\mathbf{v}^{(\varepsilon)}$ that solve (\ref{r}) and (\ref{f}) respectively. The latter is simply given by $\Delta\phi_C^{(\varepsilon)}=w_c v_C^{(\varepsilon)}$. For the former, instead, since (\ref{r}) only determines the value of the glucose import flux $v_C^{(q)}$, we searched for the simplest thermodynamically viable flux pattern among the vectors $\mathbf{v}^{(q)}$ at fixed $v_C^{(q)}$ by minimizing the $L_1$-norm \cite{de2013counting}. This effectively corresponds to performing a ``loopless'' version of FBA \cite{schellenberger2011elimination}.

Constrained Allocation FBA (CAFBA \cite{Mori2016constrained}) was instead used to compute optimal flux patterns. CAFBA is a slight but significant modification of FBA where $\mathcal{F}$ is further constrained though the additional condition described by Eq. (\ref{eq:CAFBA_constraint2}). Its implementation still only requires straightforward LP as long as the biomass composition is growth-rate independent. See \cite{Mori2016constrained} for details. To solve CAFBA, we set the costs $w_i$ of reactions in the $E$-sector to the same value, namely $w_E=8.3\times 10^{-4}$~\inversefluxunit, and used {\it E. coli}-specific values for $w_R$ and $\phi_{\max}$ as done in \cite{Mori2016constrained}. 

\subsection*{Computation of the Pareto frontier}

The Pareto front shown in Fig.~\ref{fig:ecoli_pareto}b has been computed by solving CAFBA with homogeneous costs ($w_i=w_E$ for each $i$) for different values of $w_C$, after silencing the ATP maintenance (ATPm) flux. To compensate for the lack of maintenance-associated energy costs, we increased the growth-associated ATP hydrolysis rate by an amount equal to the ATPm flux (i.e. 7.6~\mmolATPgDW \, in the iJR904 model), so that the total ATP hydrolysis flux at the maximum growth rate $\mu=1/h$ is the same as in the default model. The difference in the overall ATP hydrolysis flux (including the maintenance and growth-rate dependent component) between this implementation of CAFBA and the standard one is within 15\% for growth rates above 0.5/h. For each different class of optimal solutions, the specific intake $q$ and the specific cost $\varepsilon$ were computed, returning a set of points (one for each class) in the $(q,\varepsilon)$ plane. The Pareto front is obtained by connecting points via straight lines. Details of its construction are given in the Supporting Text. 

\subsection*{Generation of sub-optimal CAFBA solutions}
In order to generate the sub-optimal CAFBA solutions shown in Fig. \ref{fig:ecoli_pareto}b, we computed the values of $q=v_C/\mu$ and $\varepsilon=w_E\sum_i |v_i|/\mu$ for flux vectors $\mathbf{v}=\{v_c,\{v_i\}_{i\in E}\}$ different from the optimal ones. To ensure that such sub-optimal states lie sufficiently close to the Pareto front, we used flux vectors that are optimal for a version of CAFBA in which homogeneous costs $w_i = w_E$ are replaced with independent identically-distributed random variables with mean $w_E$ and dispersion $\delta$, as for the case of CAFBA with heterogeneous weights discussed in \cite{Mori2016constrained}. After obtaining a large number of such vectors for different realizations of the random costs and different values of $w_C$, we computed the corresponding metabolic state vectors $\boldsymbol{\xi}$ (by normalizing each $\boldsymbol{v}$ by its growth rate) and, for each such $\boldsymbol{\xi}$, we computed $q$ and $\varepsilon$ as defined above, i.e. using homogeneous costs $w_i=w_E$. This procedure allows to construct viable solutions that are in general sub-optimal with respect to the CAFBA solutions obtained with homogeneous costs. The depth of the sampling, that is, the typical distance of sub-optimal solutions from the Pareto front, is  controlled by the dispersion $\delta$ of the individual costs $w_i$ \cite{Mori2016constrained}. The sampled solutions approach the Pareto front as $\delta\to 0$. As $\delta$ increases, instead, the protein costs $\varepsilon$ associated to each state $\boldsymbol{\xi}$ fluctuate widely, and metabolic states far from the Pareto front become more and more likely.

\subsection*{Comparison with mass spectrometry data}

Mass spectrometry data from \cite{hui2015quantitative} include quantification of protein levels for {\it E.~coli} NQ381 (a strain with titratable lacY enzyme, based on the wild-type NCM 3722 strain) grown in minimal lactose media. Five different growth rates have been obtained by inducing different levels of lacY and, for each condition, quantitative proteomic data are available. The specific proteome cost $\varepsilon$ for the $E$-sector shown in Fig.~\ref{fig:ecoli_pareto}c was obtained in the following way. First, reactions were assigned to  the $E$-sector according to the partition used in \cite{Mori2016constrained}. Next, for each reaction, the Gene-Protein-Reaction matrix included in the iJR904 model was used to obtain a list of its corresponding enzymes. We denote by $n_{i,{\rm tot}}$ the number of reactions in which enzyme $i$ participates (irrespective of whether they are assigned to the $E$-sector or not), and by $n_{i,E}$ the number of such processes included in the $E$-sector. Given the experimental protein mass fractions $\phi_i$, our estimate for the specific $E$-sector proteome cost $\varepsilon$ is given by
\begin{equation}
 \varepsilon = \frac{1}{\mu} \sum_{i\in E} \frac{n_{i,E}}{n_{i,{\rm tot}}}\phi_i ~~.
\end{equation}
Unfortunately, the growth yields for the dataset at hand are not available. Instead, we employed an estimate obtained by solving CAFBA with heterogeneous $w_i$'s, using the default value of the ATP maintenance flux. This is justified by the quantitative accuracy that CAFBA achieves in predicting growth yields detailed in \cite{Mori2016constrained}. Notice however that the yields themselves may vary considerably across experiments. For our purposes, though, rather than the absolute value of the yield, the key is the decrease due to acetate excretion at fast growth rates, which is remarkably robust and independent on the glycolytic carbon source used \cite{Basan2015overflow}.

\bibliography{biblio_CAFBA}

\clearpage 


\section*{Supplementary material}

# A yield-cost tradeoff governs Escherichia coli's decision between fermentation and respiration in carbon-limited growth – SUPPORTING TEXT

Matteo Mori,[1] Enzo Marinari,[2,3,*] and Andrea De Martino[4,5,*]

[1]*Department of Physics, University of California, San Diego, CA, USA*
[2]*Dipartimento di Fisica, Sapienza Università di Roma, Rome, Italy*
[3]*INFN, Sezione di Roma 1, Rome, Italy*
[4]*Soft and Living Matter Lab, Institute of Nanotechnology (CNR-NANOTEC), Consiglio Nazionale delle Ricerche, Rome, Italy*
[5]*Human Genetics Foundation, Turin, Italy*

**CONTENTS**

## I. TRADE-OFFS IN THE GENERAL MODEL OF OPTIMAL PROTEIN ALLOCATION

### A. Duality of maximum growth and optimal proteome allocation

Here we consider in more detail the problem of characterizing the optimal strategy for the cell to reallocate its proteome when a specific metabolic activity is limited. In the following, $\mu$ stands for the growth rate, $v_L$ for the flux of the limited activity, $\phi_{NL}$ for the mass fraction of the rest of the proteome and $q \equiv Y^{-1} = v_L/\mu$ for the specific flux of the limited activity (per unit of growth rate). These quantities in turn depend on internal variables of the cell such as metabolic fluxes and metabolite or enzyme levels (see e.g. [1]). However, instead of considering the whole space of states spanned by such variables, we will limit ourselves to a set of cellular states ("phenotypes") which we label with an index $\alpha$. Each phenotype is assumed to be described by a different set of values for the limiting flux $v_L$ and for $\phi_{NL}$ for each growth rate $\mu$. We then define, for each $\alpha$, the quantity

$$\Phi^\alpha(w_L, \mu) \equiv w_L v_L^\alpha(\mu) + \phi_{NL}^\alpha(\mu) , \quad (1)$$

where $w_L$ denotes the cost of the protein controlling $v_L^\alpha$. Both $v_L^\alpha$ and $\phi_{NL}^\alpha$ are modulated by the dilution rate (i.e. $\Phi^\alpha \equiv$ $\Phi^\alpha(w_L, \mu)$). In particular, we make the natural assumption that both increase with $\mu$ for each $\alpha$, namely

$$\frac{dv_L^\alpha}{d\mu} > 0 \quad , \quad \frac{d\phi_{NL}^\alpha}{d\mu} > 0 \quad \forall \alpha . \quad (2)$$

For any given $w_L$, the growth rate $\mu^\alpha(w_L)$ pertaining to phenotype $\alpha$ can in principle be obtained by inverting the condition $\Phi^\alpha(\mu^\alpha, w_L) = 1$. The problem of growth rate maximization can thus be re-cast as the constrained maximization of $\mu^\alpha$ over the index $\alpha$ for given $w_L$ and $v^\alpha(\mu)$, i.e.

$$\alpha^\star = \arg\max_\alpha \mu \quad \text{subject to} \quad \Phi^\alpha(\mu, w_L) = 1 , \quad (3)$$

or, more simply,

$$\alpha^\star = \arg\max_\alpha \mu^\alpha \quad \text{subject to} \quad \Phi^\alpha(\mu, w_L) = 1 . \quad (4)$$

We call this the *direct proteome-constrained problem*. We denote by $v_L^\star$ and $\phi_{NL}^\star$ the values of $v_L$ and $\phi_{NL}$ corresponding to $\alpha^\star$. One can easily show that, for any growth rate $\mu$ for which all these quantities exist,

$$v_L^{(q)} \leq v_L^\star \leq v_L^{(\varepsilon)} \quad (5)$$

$$\phi_{NL}^{(\varepsilon)} \leq \phi_{NL}^\star \leq \phi_{NL}^{(q)} \quad (6)$$

where $(v_L^{(q)}, \phi_{NL}^{(q)})$ and $(v_L^{(\varepsilon)}, \phi_{NL}^{(\varepsilon)})$ are the values of $v_L$ and $\phi_{NL}$ respectively obtained from the solutions of

$$q\text{-problem} : \quad \alpha^{(q)} = \arg\min_\alpha v_L^\alpha(\mu) \quad \text{s.t.} \quad \mu = \mu^\alpha , \quad (7)$$

$$\varepsilon\text{-problem} : \quad \alpha^{(\varepsilon)} = \arg\min_\alpha \phi_{NL}^\alpha(\mu) \quad \text{s.t.} \quad \mu = \mu^\alpha \quad (8)$$

where the values of $\mu$ are assumed to match the optimal growth rate in the direct formulation of the problem. In order to see this, one first has to introduce the following *dual* problem of the direct proteome-constrained problem (3):

$$\alpha^\star = \arg\min_\alpha \Phi^\alpha(\mu) \quad \text{subject to} \quad \mu = \mu^\alpha . \quad (9)$$

The solution of this problem is identical to the one of the direct problem, Eq. (3), provided $d\Phi^\star/d\lambda > 0$ and the growth rate $\mu$ is set so as to match the optimal one obtained in the direct problem (see [1]). Then, the proof is straightforward: because of the definitions of the optimization problems (8) and (9), $\phi_{NL}^{(\varepsilon)} \leq \phi_{NL}^\star$, and $w_L v_L^\star + \phi_{NL}^\star \leq w_L v_L^{(\varepsilon)} \phi_{NL}^{(\varepsilon)}$, respectively. The first inequality directly gives us one of the two

---

[*] Co-last authors

constraints; another constraint, namely $v_L^\star \leq v_L^{(\varepsilon)}$, is obtained by using both inequalities. The demostration of the remaining bounds (the $q$-bounds) is analogous.

Both the $q$- and the $\varepsilon$-problem can be shown to be equivalent to two other problems different from the direct proteome-constrained problem introduced above. Indeed, the former describes the minimization of the specific flux $q = v_L/\mu$ (or to the maximization of the growth yield $Y = \mu/v_L$) at fixed growth rate, while the latter is equivalent to the maximization of the growth rate subject to a constraint on $\phi_{NL}$.

In other terms, optimal growth and optimal proteome allocation are dual to each other and solutions to the direct protein-constrained problem are bound to lie within a range defined by the $q$- and $\varepsilon$-problems. The existence of these bounds allows to study how cells may optimally handle different degrees of limitation, that is, different values of $w_L$, by switching between alternate solutions. In particular, for $w_L = 0$, the solution to the original problem coincides with the solution to the $\varepsilon$-problem. As $w_L$ increases, instead, the solution may shift towards states with larger yields (smaller specific rates) at the cost of increasing $\phi_{NL}$.

### B. Transitions between optimal phenotypes

For any given "phenotype" $\alpha$, one can obtain a growth rate $\mu^\alpha$ as a function of $w_L$ by solving the equation $\Phi^\alpha(w_L, \mu) = 1$. Optimal phenotypes are those maximizing $\mu$ for each $w_L$. Assuming for simplicity that optimal states are unique, transitions between "phenotypes" $\alpha$ and $\beta$ may occur at values of $w_L$ such that $\mu^\alpha(w_L) = \mu^\beta(w_L)$ and $d\mu^\alpha/dw_L \neq d\mu^\beta/dw_L$. Since $\Phi^\alpha = \Phi^\beta = 1$, we would have

$$w_L(v_L^\alpha - v_L^\beta) = \phi_{NL}^\beta - \phi_{NL}^\alpha \ , \tag{10}$$

i.e. any decrease in the flux $v_L$ has to be matched by an increase in the non-limited proteome $\phi_{NL}$, highlighting the tradeoff between optimal proteome allocation and efficient use of limited resources. By differentiating $\Phi^\alpha$ with respect to $w_L$ one gets

$$\frac{d\mu^\alpha}{dw_L} = -\frac{v_L^\alpha}{d\Phi^\alpha/d\mu} \ , \tag{11}$$

and therefore, assuming $d\mu^\alpha/dw_L > d\mu^\beta/dw_L$,

$$\frac{v_L^\alpha}{d\Phi^\alpha/d\mu} < \frac{v_L^\beta}{d\Phi^\beta/d\mu} \ . \tag{12}$$

This equation not only involves the absolute magnitude of the limited flux $v_L$, but also the variation of the proteome as the growth rate changes.

### C. Consequences for Constrained Allocation FBA

The consequences of equations (10) and (12) depend on the model at hand. Let us consider, as in the main text, a flux-based constraint based model with proteome allocation constraint such as CAFBA, with no maintenance ATP hydrolysis rate. In this case, as explained in the Main Text, we can introduce the metabolic states $\boldsymbol{\xi}$ and express the fluxes as $\mathbf{v} = \boldsymbol{\xi} \cdot \mu$. Let us consider a set of different metabolic states, identified by an index $\alpha$ as $\boldsymbol{\xi}^\alpha$. For each state we can compute the specific uptakes $q^\alpha$ and protein costs $\varepsilon^\star$, so that $v_L^\alpha = q^\alpha \mu$ and $\phi_{NL}^\alpha = \varepsilon^\alpha \mu$. The bounds Eq. (5) and (6) take the form

$$q^{(q)} \leq q^\star \leq q^{(\varepsilon)} \tag{13}$$

$$\varepsilon^{(\varepsilon)} \leq \varepsilon^\star \leq \varepsilon^{(q)} \ , \tag{14}$$

where $(q)$ and $(\varepsilon)$ label to specific states $\boldsymbol{\xi}^{(q)}$ and $\boldsymbol{\xi}^{(\varepsilon)}$, while the asterisk indicates the optimal state. On the other hand, Equations (10) and (12) respectively become

$$w_L(q^\alpha - q^\beta) = \varepsilon^\beta - \varepsilon^\alpha \tag{15}$$

$$\frac{q^\alpha \mu}{w_L q^\alpha + \varepsilon^\alpha} < \frac{q^\beta \mu}{w_L q^\beta + \varepsilon^\beta} \quad \Rightarrow \quad \frac{\varepsilon^\alpha}{q^\alpha} > \frac{\varepsilon^\beta}{q^\beta} \ . \tag{16}$$

Together, these two constraints imply that, at *any* transition between optimal states, both $q$ and $\varepsilon$ vary, and they do so in opposite directions. When $w_L = 0$, $\varepsilon^\star = \varepsilon^{(\varepsilon)}$, while the specific flux $q^\star$ is maximum. As $w_L$ increases, at each transition $q$ decreases as $\varepsilon$ increases. These properties lie at the basis of the Pareto front analysis.

## II. COMPUTATION OF THE PARETO FRONT

Let us consider a value $\overline{w}_C$ of $w_C$ such that states $\alpha$ and $\beta$ with $\mathcal{C}^\alpha = q^\alpha w_C + \varepsilon^\alpha$ and $\mathcal{C}^\beta = q^\beta w_C + \varepsilon^\beta$ are optimal for $w_C = \overline{w}_C$, with $\mathcal{C}^\alpha = \mathcal{C}^\beta$ when $w_C = \overline{w}_C = (\varepsilon^\alpha - \varepsilon^\beta)/(q^\beta - q^\alpha)$. We consider for definiteness $q^\alpha > q^\beta$. Starting from $w_C < \overline{w}_C$ (so with $\beta$ as the optimal state) we are interested in constraining the parameters of a sub-optimal state $\gamma$ with cost $\mathcal{C}^\gamma$, such that $\mathcal{C}^\gamma > \mathcal{C}^\alpha$ for $w_C > \overline{w}_C$ and $\mathcal{C}^\gamma > \mathcal{C}^\beta$ for $w_C < \overline{w}_C$. In what follows, we assume that $q^\beta < q^\gamma < q^\alpha$.

Suppose that $w_C < \overline{w}_C$. The constraint $\mathcal{C}^\gamma > \mathcal{C}^\beta$ can be rewritten as

$$\begin{aligned}\varepsilon^\gamma &> w_C(q^\beta - q^\gamma) + \varepsilon^\beta \\ &= w_C(q^\beta - q^\gamma) + \varepsilon^\alpha + \overline{w}_C(q^\alpha - q^\beta) \\ &> \overline{w}_C(q^\beta - q^\gamma) + \varepsilon^\alpha + \overline{w}_C(q^\alpha - q^\beta) \\ &= \varepsilon^\alpha - \overline{w}_C(q^\gamma - q^\alpha) \ , \end{aligned} \tag{17}$$

where we used $w_C(q^\beta - q^\gamma) > \overline{w}_C(q^\beta - q^\gamma)$ (since $w_C < \overline{w}_C$ and $q^\beta < q^\gamma$). Similarly, when $w_C > \overline{w}_C$, one gets

$$\epsilon^\gamma > \varepsilon^\beta - \overline{w}_C(q^\gamma - q^\beta) \ . \tag{18}$$

Note that conditions (17) and (18) identify the same half-space in the $(q^\gamma, \varepsilon^\gamma)$ plane, defined by the line passing through the points $(q^\alpha, \varepsilon^\alpha)$ and $(q^\beta, \varepsilon^\beta)$. Therefore, given a set of optimal states $\alpha, \beta, \ldots$, the Pareto frontier is obtained by connecting neighboring points with straight lines, as explained by a concrete example in Fig. 1.

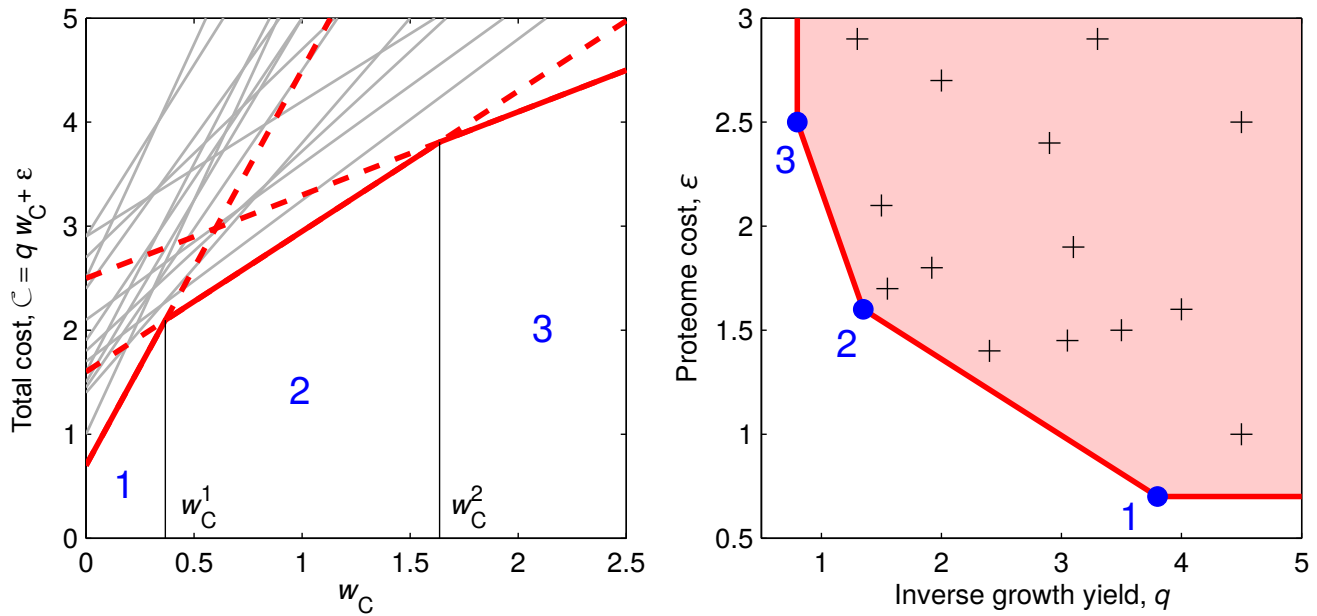

FIG. 1. Example of Pareto–optimality in CAFBA. **Left**: we show a set of possible cellular states $\alpha$, characterized by their proteome costs $\varepsilon^\alpha$ and their inverse yields $q^\alpha$. The total cost of each pathway is given by $\mathcal{C}^\alpha = q^\alpha w_C + \varepsilon^\alpha$. Each total cost $\mathcal{C}^\alpha$ is a linear function of $w_C$. The cost is minimized by the envelope (shown as a red line) of the lines corresponding to the three optimal pathways (dashed lines). The grey lines denote the costs of suboptimal pathways. Three optimal modes (carrying the lowest cost) appear as $w_C$ changes, numbered from one (minimum proteome cost $\varepsilon^1$) to three (minimum inverse yield $q^3$). The values of $w_C$ at which switches between optimal phenotypes occur are given by $w_C^1 = (\varepsilon^1 - \varepsilon^2)/(q^2 - q^1)$ and $w_C^2 = (\varepsilon^2 - \varepsilon^3)/(q^3 - q^2)$. **Right**: each grey cross in the $(q, \varepsilon)$ plane represents a suboptimal pathway. The Pareto frontier is shown in red and delimitates the feasible region. The optimal modes (blue dots) lie on the Pareto frontier.

## III. PHENOMENOLOGICAL MODEL: DEFINITION AND SOLUTION IN DIFFERENT CASES

We consider a simple reaction network shown in Fig. 2a. It involves 5 reactions among 3 chemical species, two of which ($m_1$ and $m_2$) can be exchanged with the exterior. The consumption of a "biomass" metabolite $e$ models growth. Reactions are summed up by

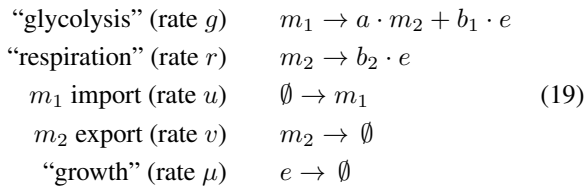

"glycolysis" (rate $g$)    $m_1 \to a \cdot m_2 + b_1 \cdot e$
"respiration" (rate $r$)    $m_2 \to b_2 \cdot e$
$m_1$ import (rate $u$)    $\emptyset \to m_1$    (19)
$m_2$ export (rate $v$)    $m_2 \to \emptyset$
"growth" (rate $\mu$)    $e \to \emptyset$

(For simplicity, the growth rate is expressed in the units of reaction fluxes.) Under steady-state mass-balance, one has $u = g$, $au = r + v$ and $\mu = b_1 g + b_2 r$. Flux states compatible with these constraints can be expressed as functions of $\mu$ and $u$ alone. From $r \geq 0$ and $v \geq 0$ one gets instead the following bounds for the growth rate,

$$b_1 u \leq \mu \leq (b_1 + ab_2)u \ , \qquad (20)$$

implying that the yield $Y = \mu/u$ of the different steady states lies between the yield of fermentation $Y_{\text{fer}} = b_1$ and that of respiration $Y_{\text{res}} = b_1 + ab_2$. As Eq. (20) does not by itself limit the growth rate, extra constraints have to be enforced to obtain well-defined solutions to the problem of maximizing $\mu$. The nature of optimal states therefore depends on which additional constraints are imposed. We consider three distinct scenarios, whose solutions are summarized in Fig. 2b–d.

**Standard FBA scenario:** A minimal choice consists in imposing an upper bound to the carbon uptake flux, $u \leq u_{\max}$, where $u_{\max}$ may coincide with the maximum carbon uptake rates observed in experiments with continuous cultures. In this condition, maximization of $\mu$ returns the state with the largest growth yield, i.e. $\mu^\star = Y_{\text{res}} u_{\max}$, $r^\star = a u_{\max}$ and $v^\star = 0$. This is a well known property of the solutions of standard FBA.

**FBA with Molecular Crowding (FBAwMC) scenario [2, 3]:**
In this case a "crowding constraint" is imposed, consisting of an overall bound on intracellular fluxes of the form $c_1 g + c_2 r = 1$. Now, for the growth rate one finds $\mu = b_2/c_2 + (b_1 - b_2 c_1/c_2)g$. For $b_1/c_1 > b_2/c_2$, the optimal solution is obtained by minimizing $g$, i.e. by setting $g = g^\star = (b_1 + ab_2)^{-1}\mu$, and presents respiratory metabolism. If instead $b_1/c_1 < b_2/c_2$, then $g$ should be maximized ($g^\star = \mu/b_1$) and the optimal solution presents fermentative metabolism. The inclusion of explicit coefficients for the carbon uptake $u$ and the fermentation flux $v$ in the crowding constraint only leads to a re-definition of the coefficients $c_1$ and $c_2$. Note however that the crowding picture for *E. coli* appears to be at odds with experiments showing that

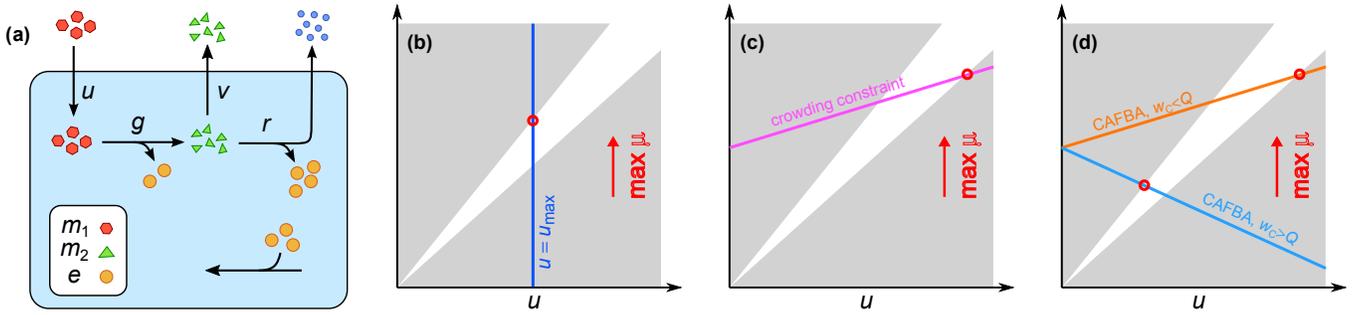

FIG. 2. **(a)** Coarse grained model of metabolism (see text). A metabolite $m_1$ can be consumed to provide "energy" $e$ plus other metabolic products $m_2$. These other products can be either further processed to yield more energy $e$ (plus waste), or excreted. The metabolites $e$ are then used by the cell to grow. **(b)-(d)** Solution space of the model imposing the steady state constraints $d[m_1]/dt = d[m_2]/dt = d[e]/dt = 0$. Grey areas denote infeasible regions. Feasible solution (white area) are delimited by the linear bounds $b_1 u \leq \mu \leq (b_1 + ab_2)u$, see Eq. (20). As $\mu$ does not have an upper limit, an extra constraint is required in order to maximize the growth rate $\mu$. We consider 3 different scenarios. **(b)** Maximization of $\mu$ subject to a maximum intake flux (standard FBA). **(c)** Maximization of $\mu$ subject to a molecular crowding constraint. **(d)** Maximization of $\mu$ subject to a proteome allocation constraint, with the lines corresponding to the cases $w_C < Q$ and $w_C > Q$ shown explicitly. Optimal solutions for each of the examples displayed are shown as red circles.

the cell density remains constant under a variety of perturbations [4, 5].

**Constrained Allocation FBA (CAFBA) scenario [6]:** In this case, the additional constraint models proteome allocation and reads $w_C u + w_1 g + w_2 r + w_R \mu = \phi_{\max}$. As in the FBAwMC case, the growth rate can be expressed as a function of $g$ alone, obtaining

$$\mu = \frac{1}{1 + b_2 w_R/w_2} \left[ \frac{b_2 \phi_{\max}}{w_2} + \frac{b_2}{w_2} (Q - w_C) g \right] \quad (21)$$

with $Q = \frac{w_2 b_1}{b_2} - w_1$. $Q$ can be interpreted as the additional protein cost of respiration, $w_2/b_2$, with respect to the cost of fermentation, $w_1/b_1$, where the protein costs are weighted by the inverse yields of the pathways. Since $w_C$ is (roughly) inversely proportional to the nutrient level, the sign of $Q - w_C$ might change when one passes from good carbon sources to poor ones. For the realistic case where $Q > 0$, in specific, the optimal solution shifts from fermentation to respiration as $w_C$ increases (i.e. as carbon is limited).



\end{document}